\documentclass{sigchi}

\usepackage{balance}  
\usepackage{graphics} 
\usepackage{times}    
\usepackage{url}      

\makeatletter
\def\url@leostyle{%
  \@ifundefined{selectfont}{\def\UrlFont{\sf}}{\def\UrlFont{\small\bf\ttfamily}}}
\makeatother
\def\pprw{8.5in}
\def\pprh{11in}

\usepackage[pdftex]{hyperref}
\hypersetup{
pdftitle={SIGCHI Conference Proceedings Format},
pdfauthor={LaTeX},
pdfkeywords={SIGCHI, proceedings, archival format},
bookmarksnumbered,
pdfstartview={FitH},
colorlinks,
citecolor=black,
filecolor=black,
linkcolor=black,
urlcolor=black,
breaklinks=true,
}

\usepackage{graphicx}
\usepackage{balance}
\usepackage{comment}
\usepackage{times,listings,amsfonts}
\usepackage{amsmath,xcolor}
\usepackage{graphicx}
\usepackage{pifont}
\usepackage{xspace}
\usepackage{url}

\begin{document}
\vspace{-0.8cm}

\title{Managing Commercial HVAC Systems: \\What do Building Operators Really Need?}
\numberofauthors{1}
\author{
	\alignauthor Bharathan Balaji$^\dag$, Nadir Weibel$^\dag$, Yuvraj Agarwal$^\ddag$\\ 
	\vspace{2mm} 
	\affaddr{$^\dag$University of California, San Diego \hspace{12mm} $^\ddag$Carnegie Mellon University}\\
	\affaddr{\hspace{10mm}$^\dag$San Diego, USA \hspace{35mm} $^\ddag$Pittsburgh, USA}\\
	\email{$^\dag$\{bbalaji, weibel\}@ucsd.edu\hspace{22mm}$^\ddag$yuvraj.agarwal@cs.cmu.edu}\\
}

\maketitle

\begin{abstract}
Buildings form an essential part of modern life; people spend a significant amount of their time in them, and they consume large amounts of energy. A variety of systems provide services such as lighting, air conditioning and security which are managed using Building Management Systems (BMS) by building operators. 
To better understand the capability of current BMS and characterize common practices of building operators, we investigated their use across five institutions in the US. We interviewed ten operators and discovered that BMS do not address a number of key concerns for the management of buildings. Our analysis is rooted in the everyday work of building operators and highlights a number of design suggestions to help improve the user experience and management of BMS, ultimately leading to improvements in productivity, as well as buildings comfort and energy efficiency.
\end{abstract}

\vspace{-0.5cm}
\category{H.5.m.}{Information Interfaces and Presentation (e.g. HCI)}{}

\section{Introduction}
Buildings are central to our life with people spending an average of 87\% of their time inside them~\cite{klepeis2001national}. Buildings also contribute to 39\% of the energy usage in the USA~\cite{doe2011buildings}. Modern buildings consist of various types of systems to meet the requirements of the occupants. As digital technology gets integrated into buildings, humans increasingly interact with these systems using computer-based interfaces. 

\begin{figure}[b!]
\centering
\includegraphics[width=1.0\linewidth]{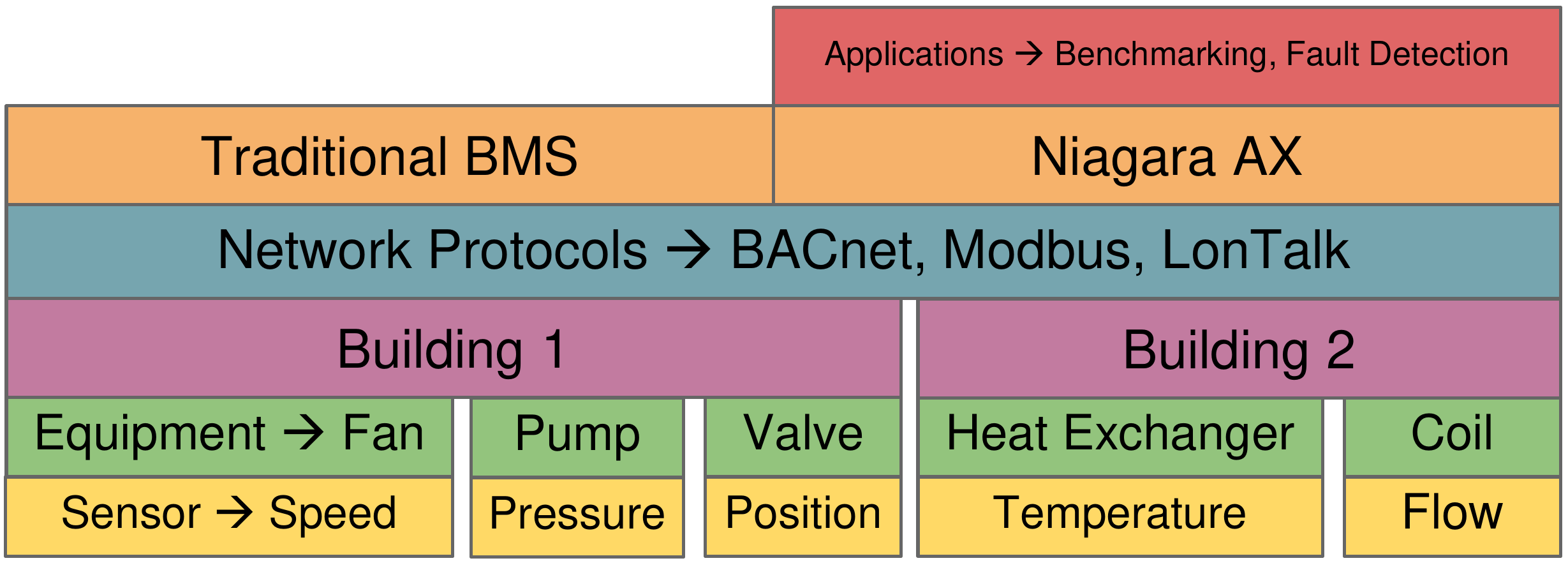}
\caption{Architecture of a typical Building Management System}
\label{fig:bms_arch}
\end{figure}

We focus on commercial buildings, which constitute roughly half the total building-related energy usage. These buildings make use of Building Management Systems (BMS) to monitor, operate and maintain the various services within them. BMS consist of networked sensors and actuators operated using a centralized server hosting the BMS UI. 
Typically, building occupants do not interact with BMS directly, but use regular switches/thermostats, and send their complaints to building managers. Figure~\ref{fig:bms_arch} outlines BMS's main architecture. BMS are used for management of Heating, Ventilation and Air Conditioning (HVAC), lighting, security, irrigation, etc., and these systems are usually not integrated. 
They consist of complex set of equipment and control programs used by a few key operators such as building managers, maintenance personnel or service contractors. As large equipment can be controlled using BMS software, even small actions can affect the comfort and energy efficiency of the entire building.

We studied the usage of BMS across five institutions in the US and outline here the challenges of everyday use of these systems. Our analysis shows that these challenges are often due to incorrect design and development of the current systems, and we suggest design changes to help overcome them. Using contextual inquiry~\cite{beyer1997contextual}, we interviewed participants with diverse designations who interact with BMS software regularly. We focus on HVAC management using BMS as it is the only system installed in most buildings, and the software has advanced to address the challenges that emerge in a complex system. Our contribution is therefore two-fold: first we contextually document key challenges of BMS use across operator roles, BMS software, type of institution, and geographic location; second, we distill important insights and design directions that can be incorporated in the development of the next generation interfaces.

%
%





\begin{table*}[t!]
\resizebox{\textwidth}{!}{
\begin{tabular}{p{5cm}p{6cm}p{7cm}p{2cm}}
\multicolumn{1}{c}{\textbf{Institution/Company}} & \multicolumn{1}{c}{\textbf{Participants}} & \multicolumn{1}{c}{\textbf{BMS}} & \multicolumn{1}{c}{\textbf{Buildings}}\\[0.1cm]\hline\\[-0.1cm]
University of California, San Diego	  & \parbox{6cm}{Energy manager (P1), HVAC technician~(P2), Building Manager (P3) } & \parbox{7cm}{Johnson Controls Metasys} & \parbox{2cm}{\centering100+}\\[0.3cm]                                                                                                                                                                     
University of San Diego   & \parbox{6cm}{HVAC technician (P4)} & \parbox{7cm}{Siemens Apogee} & \parbox{2cm}{\centering 50+}\\[0.1cm]                                                                                                                                                                  
Carnegie Mellon University   & \parbox{6cm}{Asset Preservation Manager (P5)} & \parbox{7cm}{Automated Control Logic, Automatrix, Siemens, Johnson Controls} & \parbox{2cm}{\centering 100+}\\[0.3cm]                                                                                                                                                                  
University of California, Berkeley   & \parbox{6cm}{Two building managers (P6, P7)} & \parbox{7cm}{Automated Control Logic, Siemens, Barrington} & \parbox{2cm}{\centering 100+}\\[0.1cm]                                                                                                                                                                  
San Diego County  & \parbox{6cm}{HVAC maintenance operator (P8)} & \parbox{7cm}{Tridium Niagara AX} & \parbox{2cm}{\centering 520+}\\[0.1cm]                                                                                                                                                                  
Johnson Controls Inc.   & \parbox{6cm}{BMS technician (P9)} & \parbox{7cm}{Johnson Controls Metasys} & \parbox{2cm}{\centering -}\\[0.1cm]                                                                                                                                                                  
Enernoc Inc.   & \parbox{6cm}{Energy efficiency consultant (P10)} & \parbox{7cm}{Various BMS, Enernoc Insight} & \parbox{2cm}{\centering -}\\[0.1cm] \hline                                                                                                                                                                 
\end{tabular}
}
    \caption{Portfolio of participants in our user study, the BMS platform they used and the number of buildings managed by the institution} 
    \label{tab:participants}
\end{table*}

\section{Motivation, Background and Related Work}
Several stakeholders are involved in the management of HVAC systems. BMS operators can be grouped into three categories: people who ensure (a) day to day operation, (b) comfort of the occupants, and (c) energy efficiency. Facility managers and operators are common to all buildings in an institution. Building managers work at the building level, and help maintain all systems in the building. Issues which require technical expertise are forwarded to the facilities management. In addition, service contractors who specialize in certain services conduct repairs, upgrade software, etc. Finally, commissioning consultants are hired short term to ensure that building systems are functioning correctly and recommend changes to equipment or control programs. We use the term ``\emph{operator}'' to refer to the different users of BMS.


Despite the need to employ different professionals for managing buildings, institutions cut costs by having less staff members, relying primarily on the BMS to assist with \mbox{monitoring} and automation of HVAC systems. BMS are routinely installed in newly constructed buildings, and older buildings are upgraded to BMS-enabled equipment. BMS provide services such as monitoring of sensor data across the system, programming of control sequences for proper equipment operation, reporting faults detected through sensor data, graphical visualization, and providing access control across users. 

Traditionally, BMS are provided by HVAC equipment vendors such as Johnson Controls, Siemens, and Automated Logic, as end to end customer solutions. Although proprietary solutions typically do not communicate to systems from other vendors, communication protocols such as BACnet and Lontalk were introduced to increase interoperability across vendors. Nevertheless, compatibility remains a challenge as each vendor uses their own extensions of common protocols. 

To overcome a number of those interoperability challenges, Tridium introduced Niagara AX,\footnote{http://www.niagaraax.com} a vendor-agnostic BMS platform that provides interoperability across different vendors as well as support for development of third party applications. Niagara AX is based on BAJA (Building Automation Java Architecture~\cite{baja}), supports upcoming standards such as oBIX and Haystack, and is becoming the de-facto standard for vendor-agnostic BMS implementation. 

Despite the efforts so far to improve BMS, the work of operators remains challenging, and is becoming increasingly complex. With hundreds of buildings each with thousands of sensors, the design of effective, efficient, and satisfactory BMS is key to overcome challenges and avoid operator overload. We report on a variety of issues that operators face with today's systems and outline directions for the next generation BMS. 





\begin{table*}[t!]
\resizebox{\textwidth}{!}{
\begin{tabular}{p{5cm}p{16.5cm}}
\multicolumn{1}{c}{} & \multicolumn{1}{c}{}\\[0.2cm]\hline\\[-0.2cm]
\textbf{Contextual and Historical Data} & \parbox{16.5cm}{Third party products for search and visualization, including historical sensor data, can be added on top of Niagara AX. Operators can have personalized dashboards that show relevant indicators such as jumps in energy consumption.}\\[0.35cm]                                                                                                                                                                     
\textbf{Naming Convention} & \parbox{16.5cm}{Introduced component object model for naming building entities -- sensors, sensor metadata, actuators, and control sequences. Supports rising standards such as oBIX and Haystack.}\\[0.35cm]                                                                                                                                                                     
\textbf{Fault Reporting}                         & \parbox{16.5cm}{SkySpark is a popular third-party tool for analyzing HVAC sensor data for fault detection and diagnosis that can be installed on top of Niagara AX. It supports open standards, and provides relevant information on each fault to the user. }\\[0.35cm]                                                          

\textbf{Data Analysis} & \parbox{16.5cm}{The platform supports storing of sensor data and auditing of user actions to understand the historical performance of the HVAC. Historical data also allows users to easily benchmark performance with respect to their past data.}\\[0.35cm]                                                                                                                                                                     
\textbf{Vendor Lock-in} & \parbox{16.5cm}{JACE box is an intermediary between vendor equipment and the Niagara AX. Boxes contain drivers to port vendor specific protocols to proprietary protocols. Data is exposed to third party applications using BAJA open standard.}\\[0.35cm]                                                                                                                                                                     
\textbf{Search and Reporting} & \parbox{16.5cm}{Third party applications developed in Java or as a web service enables data querying using SQL-like language. Reports can be built to periodically summarize performance, usage and energy characteristics.}\\[0.2cm]\hline                                                                                                                                                                     
\end{tabular}
}
\caption{Innovations and improvements introduced by Niagara AX with respect to the building operators needs outlined in the interviews.} 
\label{tab:niagaraax_innovations} 
\end{table*}

\section{Understanding Building Operators}
To understand the current experiences of building operators, we studied the use of BMS by ten building operators across five institutions managing more than 870 buildings (Table~\ref{tab:participants}). We followed a hybrid semi-structured~\cite{louise1994collecting} and contextual learning~\cite{beyer1997contextual} model that elicited direct feedback from the users and engaged them in detailed description of their experiences. We conducted all but two interviews on-site, at the participants office or in a nearby conference room, with the remaining two conducted remotely. 
We took detailed notes during all interviews, and with participants consent recorded audio for 7 out of 10 interviews, and for five of them we collected videos of the operator's interaction with the system. We transcribed the recorded interviews for in depth analysis. 

Data collected was analyzed by two researchers who worked in the area of smart buildings for five years and an expert in human-computer interaction and user interfaces. We exploited elements from grounded theory~\cite{strauss1990basics} to perform a thematic analysis and we grouped emerging elements into seven key challenges that building operators currently face.

\subsection{Challenges in Building Management}
Regular maintenance of buildings include addressing comfort complaints, resolving BMS alarms indicating faults in building system, performing periodic tasks such as replacement of dirty filters or installing/upgrading of equipment or software. 


All of our participants felt they were understaffed and underfunded to handle the number of issues they need to address as summarized by P10 (consultant): ``\emph{Almost everywhere we go they don't have enough maintenance staff to do things right most of the time. So that's very common that they do the quick fix rather than the right fix}''. Operators admitted that they were aware that many of their buildings are operating inefficiently, but did not know which ones were inefficient and how inefficient they were. Their main strategies to overcome this issue were based on their past experiences, and on the age of equipment. They also relied on commissioning -- i.e. manual checks of specific buildings -- to identify major inefficiencies. Although sometimes effective, these strategies are not scalable, do not transfer well to other operators, are not documented, and are not sustainable for a large campus. 

To better understand the underlying reasons behind these challenges we now detail seven key problems that we identified across our interviews:

\textbf{Simplistic Fault Reporting:} The way faults are reported to the operators are simplistic and create an alert every time a sensor value goes beyond a pre-specified threshold. The underlying cause of the alert is not easily identifiable: it could be related to a sensor drift, an error in configuration, a damage in equipment or a combination of factors. Alerts which are related to each other are not grouped together, causing a deluge of alerts for the same fault. Therefore, faults often accumulate and some of them remain unresolved. For instance, one of the operator showed us $>$100,000 alerts that accumulated in her system that she would never be able to catch up with. Also, although energy efficiency is increasingly important, sensors installed only target critical faults to reduce costs.

\textbf{Missing Contextual and Historical Data:} Analyzing the underlying cause of a fault is vital to locate the problem and fix it. To this extent, various levels of data needs to be available to operators to analyze the status of the system. While historical sensor data was provided in all BMS, in one of the universities data was only stored for 3 days, and in another the trending had to be started manually, which at times happened only after the discovery of a fault. P1 expresses his frustration: ``\emph{one of its [data trending] biggest limitations is that I'm always being told that \textquoteleft{Don't ask us to map so many points}\textquoteright or \textquoteleft{Don't ask us to set too many trends}\textquoteright  because it'll overload the system.}'' Furthermore, relevant information is distributed across a variety of sources. Contextual information such as equipment location, connection to other units, model number, etc. are not available. In one institution, the power meter data is accessed separately from the HVAC sensor data, and the relationship between different equipment is only present in architectural drawings. This missing information results in the operator visiting the site in person to diagnose the fault, which increases the time to fix it considerably. 

\textbf{Inconsistent Naming Convention:} Even within the same institution and BMS, we witnessed lack of standard naming conventions across different buildings. Names are manually labeled by different operators (and, even renamed over the years), and therefore, do not follow a consistent naming convention. As P10 explains: ``\emph{I don't know if they do anything to make their point names consistent [...] sometimes they'll just leave them as AV1. And that's not very helpful to anybody.}'' 

\textbf{No Integrated Data Analysis Tools:} BMS only provided raw sensor data, and did not support easy addition of computed information. Thus, participants reported having to perform many calculations by hand to analyze data. As P1 explains: ``\emph{I was looking at that specific room [...] to see things like \textquoteleft{Ok does the total supply flow match up to the total exhaust flow?}\textquoteright I was doing summations in my head of these numbers, like ... is this making sense?}''. BMS do not provide common metrics to be used with analysis such as benchmark against other buildings or calculate efficiency of operation.

\textbf{Vendor Lock-in:} Traditional BMS lock-in the facilities with their equipment, so operators can only use vendor-provided hardware and software. Even when open protocols such as BACnet are adopted, vendor extensions of the protocol do not match with other vendors. These vendors also provide versions of Niagara AX platform which are incompatible with other vendors. As P8, who uses a Niagara system complains: ``\emph{I wish that there was a way that I can put a third party item onto it so I don't have to upgrade the whole system. But that's not available. It has to come as a part of what they sell.}''


\textbf{Forgotten Overrides:} When fixing certain faults, it is common for operators to temporarily override current settings to conduct repairs. However, operators frequently forget to release their override leading to faults. BMS support storing past operator operations, but the number of entries is limited. No option is provided to integrate notes while overriding settings to enable later analysis. As P10 recalls from experience: ``\emph{...~extremely common would be operators leaving something in override. So either switching the handoff auto switch on a VFD by hand and just leaving something flat out. Typically these are things that I'll do with the intention of having it be temporary but then you're too busy to come back and fix the root of the problem so it just stays for weeks or years.}''

\textbf{Recognize Occupants Misuse:} Occupants cause faults because they are unaware of how HVAC systems work. Space heaters are commonly used during winter, causing excessive energy wastage. Refrigerators or other appliances block thermostats or air vents, causing incorrect operation of HVAC. It is difficult to understand and recognize from the BMS when a faulty operation is due to a misuse: ``\emph{... a pet peeve of mine is when people set their air conditioner to like 69 degrees and the [occupants] have the 1500 watt electric heaters going on their desks at the same time.}'' [P4].

\section{Discussion}
As outlined above, facility managers struggle with integration of different systems, lack of standardized data formats and are locked into vendors after the initial installation. The infrastructure for historical data collection is not robust, contextual information that is key to understanding the underlying situation is missing, and data analytics are simplistic, putting the onus on the operator to do calculations.







Addressing those issues through vendor-specific HVAC systems is hard, since they are monolithic and have not been designed with flexibility in mind. However, the vendor-agnostic solution provided by Niagara AX has the potential to overcome some of these limitations (Table \ref{tab:niagaraax_innovations}). One of the institutions in our study had Niagara AX installed in eight (out of 520) buildings. The building manager reported that BMS is easier to use and helped with benchmarks and faults. 

Despite the functionality introduced by Niagara AX, only one of the operators we interviewed makes use of it. This is due to several reasons. First, many of our interviewees were unaware of the benefits of vendor-agnostic platforms such as Niagara AX. Second, even though it helps in the long-term, installation costs in existing buildings are non-trivial: much of the costs account for manual translation of data from existing building(s) to Niagara platform. 
Finally, changes to another system result in at least a temporary drop in productivity and will introduce a variety of new and different interfaces that operators are not willing to embrace easily.

\subsection{Designing Next Generation Interfaces}
While the approach put forward by Niagara AX and the overall idea of vendor-agnostic BMS is useful, more radical changes are needed to exploit the inherent energy saving capability and improve occupant comfort in buildings. Based on the outcome of the interviews we identify a finite but carefully investigated list of design recommendations next.

\textbf{Automation:} Several parts of BMS remain expensive to create or program because of lack of automation and use of standard machine readable formats. Graphics for floor plans and equipment connections, for example, are hand drawn. Building architectural and mechanical plans are available in CAD drawings with standard formats like Green Building XML, and should be leveraged by BMS. Similar automation features can be developed for discovery of sensors installed, population of metadata such as location, and acquisition of equipment datasheets by providing them in machine readable form.

\textbf{Data Analytics:} With the amount of data available from sensors in the HVAC system, a wide variety of analytics and diagnostics can be performed. However, operators in our study stressed how the current system: ``\emph{is definitely suboptimal right now, I'm overwhelmed by the amount of data that's available and the lack of automation of it.}'' [P1]. Several tools and techniques have been created by researchers~\cite{crawley2001energyplus,katipamula2005review}
to develop HVAC system models, detect faults and inefficiencies, and simulate different scenarios. However, these tools remain disconnected from BMS. We believe that integrating them into the BMS will provide useful insights to operators.

\textbf{Contextual Information:} When operators make changes to the HVAC system (e.g. temperature control), they do not get feedback on the effects on energy or comfort. Instead, they judge these effects based on the raw values of sensors provided, although effects are only visible after several minutes to hours. Operators need immediate feedback using form and metrics that is relevant 
(e.g. type of room or energy consumed). It will help them be more efficient and reduce mistakes. P1 commented on this point stating how ``\emph{If operators were getting immediate feedback in terms of energy waste ... as soon as you put that override in it could be like \textquoteleft{This is going to cost the university \$40,000 this year, are you sure?}\textquoteright}''


\textbf{Communication Among Stakeholders:} Communication among management personnel is done through phone calls, emails and work orders, all of them kept separate from BMS. Novel BMS need to support contextual annotations to avoid misunderstanding. For example, the energy manager can infer that an override is in place due to repairs being conducted. BMS also need to involve occupants as their actions directly impact the system. An operator, for instance, can see that HVAC is active late at night due to a special occassion.

\textbf{User Support:} BMS usability would greatly increase by adopting standard user involvement practices. Instead of requiring specialized training, operators should be provided with wiki pages and discussion boards to encourage learning and adoption of best practices. BMS needs to support varying requirement of different roles in management, and provide appropriate levels of abstraction, permissions and data analytics to assist them in their daily work.





\section{Summary and Conclusion}
We collected data across five large institutions in the USA and engaged in discussions with ten different building operators on their experiences, frustrations and their informed requirements for novel BMS. After analyzing hours of interviews, we identified seven key recurrent problems that need to be addressed. With the help of our participants' suggestions we then distilled a list of overall directions to take for the design of the next generation BMS. We believe that these results have direct applicability and can be used to guide the development of novel interfaces, like the one that we are currently designing as part of our larger research on building management. We hope that the depth and breadth of our findings will support the much needed change in perspective and the bootstrapping of a new class of flexible, user-centered BMS.

%
%
%
%
%
\balance


\bibliographystyle{acm-sigchi}
\bibliography{CHI2015_HVAC}
\end{document}